\documentstyle[sprocl]{article}

\input{psfig.sty}

\def\be{\begin{equation}}
\def\ee{\end{equation}}
\def\bea{\begin{eqnarray}}
\def\eea{\end{eqnarray}}

\newcommand{\bk}{{\bf k}}

\begin{document}

\title{DIFFRACTIVE VECTOR MESON PRODUCTION IN A UNIFIED
 $\kappa$-FACTORIZATION APPROACH}

\author{I.P. IVANOV, N.N. NIKOLAEV}

\address{IKP (Theorie) Forschungszentrum J\"ulich, Germany\\
E-mails: i.ivanov@fz-juelich.de, n.nikolaev@fz-juelich.de}

\maketitle \abstracts{In the framework of the
$\kappa$-factorization approach and on the basis of recently
determined DGSF, we calculated the production rates of ground and
excited ($2S$ and $D$ wave) states of vector mesons and
investigated their various kinematical and spin dependencies.
We also addressed the issue of $S/D$ wave mixing in quarkonia.}

\section{The vector meson production amplitude}
The general amplitude of diffractive production
of vector meson $V$ is well known: the basic quantity
is the cross section of $q\bar q$ color dipole interaction
with the proton, which is convoluted with the initial photon
and final vector meson wave function (WF). Algebraically,
the amplitude in the diagrammatically straightforward
$\kappa$-factorization approach reads:
\begin{eqnarray}
{\cal A}(x,Q^{2},\vec \Delta)=
is{c_{V}\sqrt{4\pi\alpha_{em}}
\over 4\pi^{2}}
\int_{0}^{1} {dz\over z(1-z)} \int d^2 \vec k  \psi(z,\vec k)
\nonumber\\
\cdot\int {d^{2} \vec \kappa \over
\vec\kappa^{4}}\alpha_{S}(q^2)
\left(1 + i {\pi \over 2} {\partial\over \partial\log x}\right)
{\cal{F}}(x,\vec \kappa,\vec \Delta)
\cdot I(\gamma^{*}\to V)\nonumber
\end{eqnarray}
This expression contains two pieces non-calculable within pQCD:
the vector meson WF $\psi(z,\vec k)$ and the differential gluon
structure function (DGSF) ${\cal{F}}(x,\vec \kappa,\vec \Delta)$.
The principal novelty of this work is that DGSF is now under
control. The off-forward DGSF ${\cal{F}}(x,\vec \kappa,\vec
\Delta)$ can be linked to the forward DGSF, which
has been recently determined from experimental data on
$F_{2p}$ in the whole $Q^2$ domain \cite{main}. This dramatically
reduces the level of ambiguity in diffractive vector meson
production calculations, leaving us only with one unknown
quantity --- the vector meson WF.

The vector meson WF involves two components: the spinorial
structure of the $q\bar q V$ vertex and the radial WF
$\psi(\bk^2)$. In our work we consistently used the spinorial
structures corresponding to $q \bar q$ pair sitting in pure $S$ or
$D$ wave state \cite{waves}. This allows us to address production
of both ground ($1S$) and excited ($2S$, $D$ wave) vector meson
states as well as $S/D$ wave mixed states. For the radial WF we
chose a simple soft WF Ansatz (i.e. without specific large-$\bk$,
or short distance, enhancement) with parameters adjusted to match
the vector meson $V\to e^+e^-$ decay width.

\section{Production of ground states}
We calculated the production rates of ground ( $\equiv 1S$) state
vector mesons $\rho^0, J/\psi, \Upsilon(1S)$ and compared
predictions with the experimental data. We observed good agreement
in a broad $Q^2$ range and confirmed the scanning phenomenon (the
cross sections of different vector mesons, corrected by the flavor
factor, are almost the same when expressed in terms of the scaling
variable $Q^2+m_V^2$). We also observed that the predicted energy
dependence and $|t|$-dependence are the same as inferred from the
experiment (this is best illustrated by plotting the effective
intercepts and effective slopes as functions of the same scaling
variable).

We also compared cross sections $\sigma_L$ and $\sigma_T$ in the
$\rho^0$ meson case with experiment. We found good description of
$\sigma_L$ but observed a strong systematic departure of our
predictions for $\sigma_T$ from experimental data in the
high-$Q^2$ region: our curves go substantially lower than the
data. The disagreement between predicted $\sigma_L/\sigma_T$ ratio
and the data is, of course, the consequence of this $\sigma_T$
puzzle. Since the gluon density is under control, such a deviation
can only be attributed to the WF Ansatz. In other words, it seems
that the soft WF does not exhaust the whole physics at short
distances. Another possible solution will be given later.

Since in our approach we took into account all helicity
amplitudes, both $s$-channel helicity conserving and violating
ones, we have predictions for the whole set of density matrix
elements. We observed that our predictions for $\rho$ and
$\omega$ mesons are in agreement with experiment.

\section{Production of excited states}
We calculated the cross section of $2S$ and $D$ wave vector meson
production in the case of $\rho$-system and charmonium as
functions of $Q^2$. These cross sections were found suppressed in
comparison with $1S$ states, the magnitude of suppression
following remarkably different $Q^2$ behavior for $2S$ and $D$
wave states. This difference is due to distinct mechanisms
of suppression: in the case of $2S$ state this is the well-known
node effect, while for $D$ wave state it comes from vanishing
radial WF at origin.

For illustrative purposes we also calculated ratios
$\sigma_L/\sigma_T$ and density matrices for production of $2S$ and
$D$ wave states in the $\rho$ system. The prominent features of
$\sigma_L/\sigma_T$ (Fig.1) (non-monotonous $Q^2$ behavior for $2S$ state
and strong, about one order of magnitude, suppression for $D$
wave state) should serve as very clear experimental indicators in
extracting excited vector meson signal. Also, many of density
matrix elements, especially helicity violating ones, are very
distinct in the case of $1S/2S/D$ wave states.

\begin{center}
\begin{figure}[t]
\hspace{3cm}
\psfig{figure=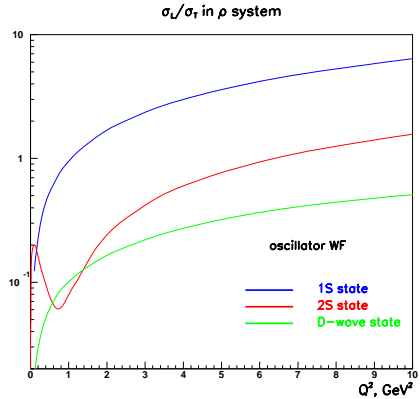,height=2.5in}
\caption{Ratio $\sigma_L/\sigma_T$ for $1S/2S/D$ states
in the $\rho$-system \label{fig:siglt}}
\end{figure}
\end{center}

\section{$S/D$ wave mixing}
It is known that tensor forces, which naturally appear in quark
potential models, lead to $S/D$ wave mixing of $q\bar q$ states.
This mixing can be quite strong, which is indicated by the
$e^+e^-$ decay width of $D$-wave candidate in the charmonium
spectrum $\psi(3770)$. As an example of effects that $S/D$ mixing
can lead to we calculated ratio $\sigma_L/\sigma_T$ for $\rho$
meson in the presence of mixing. We saw that the above mentioned
$\sigma_L/\sigma_T$ puzzle can be, in principle, eliminated at the
expense of strong $S/D$ wave mixing. Further work is necessary to
find out if this is the case.


\section*{References}
\begin{thebibliography}{99}
\bibitem{main}I.P.~Ivanov, N.N.~Nikolaev, E-print:
hep-ph/0004206 (2000); I.P.~Ivanov,
{\em "Anatomy Of The Proton Structure Functions
In $\kappa$-Factorization"}, these proceedings.

\bibitem{waves}I.P.~Ivanov, N.N.~Nikolaev, {\em JETF Lett.}
{\bf 69} (1999) 294; I.P.~Ivanov, MSc thesis,
E-print: hep-ph/9909394 (1999).


\end{thebibliography}
\end{document}